\documentclass[twocolumn,showpacs,preprintnumbers]{revtex4-1}
\usepackage[dvips]{graphicx}      
\usepackage{dcolumn}     
\usepackage{bm}
 
\begin{document}           

\title{Metamagnetism and Weak Ferromagnetism in Nickel (II) oxalate crystals}       
\author{E Romero$^1$, M E Mendoza$^1$, and R Escudero$^2$}
\address{$^1$Instituto de F\'{\i}sica, Benem\'{e}rita Universidad Aut\'{o}noma de Puebla, Apartado Postal J-48, Puebla 72570, M\'{e}xico}
\address{$^2$Instituto de Investigaciones en Materiales, Universidad Nacional Aut\'{o}noma de M\'{e}xico, Apartado Postal 70-360,  M\'{e}xico DF 04510, M\'{e}xico} 

\email[Author to whom correspondence should be addressed. ER, email address:] {evarotel@gmail.com}

\date{\today }  

\begin{abstract} 

Microcrystals of orthorhombic nickel (II) oxalate dihydrate were synthesized through a precipitation reaction of 
aqueous solutions of nickel chloride
and oxalic acid. Magnetic susceptibility exhibits a sharp peak at 3.3 K and a broadrounded maximum near 43 K. 
We associated the lower maximum with a metamagnetic transition that occurs when the magnetic field is about $\geq$ 3.5 T. 
The maximum at 43 K is the typical of 1D antiferromagnets, whereas weak ferromagnetism behavior was observed in the range of 3.3 to 43 K.

\end{abstract}      

\maketitle 

\section{Introduction}  

It has been shown, theoretically and experimentally, that if a critical
magnetic field H is applied along the crystalline anisotropy axis of an
antiferromagnet (AF),  there occurs a nearly $90^{o}$ rotation of the sublattice
vectors. An antiferromagnetic material with such a beha\-vior is known as a
metamagnet \cite{Keffer-73,Jones-72}. This behavior  has been observed in
compounds like: $CuCl_{2}\cdot 2H_{2}O$ \cite{Liu-06}, $FePt_{3}$, $YCO$, $
TiBe_{2}$ \cite{Hurst-82}, and $Ni(C_{2}O_{4})(bpy)$, where $bpy=4,4^{\prime
}-bipyridine(C_{10}H_{8}N_{2})$ \cite{Yuen-06}, when  applying high magnetic
fields (from 10 to 50 T). The type of ordering in $Fe(C_{2}O_{4})(bpy)$,
and $Co(C_{2}O_{4})(bpy)$ is also antiferromagnetic but with canted spins. 
This uncompensated antife\-rromagnetism  produces  weak ferromagnetism (WF). Recently, investigations in the
quasi-one-dimensional magnetic compound $\beta -CoC_{2}O_{4}\cdot 2H_{2}O$ 
\cite{Romero-11} revealed that the intra and interchain magnetic
interactions tilt the spins, distorting the antiferromagnetic order and
producing the WF.  

The orthorhombic $\beta -$phase of nickel oxalate
dihydrate belongs to the space group $Cccm$ with cell parameters $a=11.842(2)$ \AA ,
$b=5.345(1)$ \AA , and $c=15.716(2)$ \AA \cite{Deyrieux}.
In this structure, oxalate bridges link $Ni^{2+}$ ions chains along the $b$
direction \cite{Dubernat,Molinier,Deyrieux69,Druet-99} and they mediate a
dominant antiferromagnetic intrachain superexchange interaction \cite%
{Kurmoo-09}. Its molar susceptibility data exhibits the broad rounded maximum
typical of 1D antiferromagnets with $T_{\max }\sim 41$ K \cite{Vaidya08}.   

In this paper we report results on the synthesis, crystal structure, and
magnetic measurements of the orthorhombic $\beta -NiC_{2}O_{4}\cdot 2H_{2}O$
microcrystals. We determined by isothermal $M-H$ measurements, from 2 to 80 K, 
that this phase exhibits weak ferromagnetism in the range of 3.3 to 43 K. At low temperature a metamagnetic order 
is observed under an applied field $\geq $ 3.5 T, in $\chi -T$ measurements.

\section{Experimental Methods}  

The synthesis of $\beta$ $-$ nickel oxalate dihydrate under nitrogen
atmosphere was carried out by precipitation reaction of aqueous solutions
of nickel (II), chloride 0.1 M (Aesar, $99.9995\%$) and oxalic acid 0.00625 M (Baker, $\geq99.9\%$), following the chemical equation,

\begin{equation}
H_{2}C_{2}O_{4(aq)}+NiCl_{2(aq)}\rightarrow NiC_{2}O_{4}\text{\textperiodcentered}2H_{2}O_{(s)}+2HCl_{(aq)}.   
\end{equation}

The precipitates were filtered and dried at room temperature. Morphological analyses were performed with a scanning electron microscope
(SEM) Cambrige-Leica Stereoscan 440. Powder X-ray diffraction patterns
were acquired using a Siemens D5000 diffractometer opera\-ting in the
Bragg-Brentano geometry with $\lambda $(Cu-K$\alpha $) = 1.541 \AA , and 
2$\theta $ scan from 10 - 70$^{\circ }$, and step size of 0.02$^{\circ }$.
Thermogravimetric (TG) and differential thermal analy\-sis (DTA) curves were
obtained in a SDT-TA Ins\-truments model 2960 in air atmosphere with a
heating rate of 5 $^{\circ }$C/min, from room temperature up to 600 $^{\circ
}$C. Magnetization measurements were done with a Quantum Design MPMS SQUID
magnetometer, MPMS-5. Zero Field Cooling (ZFC) and Field Cooling (FC) cycles
were performed at magnetic field intensities from 0.01 T to 5.00 T, in
the range 2 up to 250 K. Isothermal magnetization measurements $M(H)$
were obtained from 2 K - 80 K. The diamagnetic contribution calculated
from Pascal's constants \cite{Bain08} was $\chi _{Di}=-72\cdot 10^{-6}$ cm$
^{3}$/mol.

\section{Results and Discussion}    

\subsection{Synthesis} 

Prismatic green microcrystals of nickel oxalate dihydrate were obtained
after eight days of growth in nitrogen atmosphere. Figure 1
presents a SEM micrograph of the crystals with average length size (L) and
diameter (D) of 1.47 and 0.65 $\mu$m, respectively. Chemical
analysis by digestion/ICP-OEP, combustion/TCD/IR-detection and
Pyrolysis/IR-detection, gave the composition in weight of 31.82 \% in
nickel, 13.21 \% in carbon, 2.48 \% in hydrogen and 51.00 \% in
oxygen.

\begin{figure}[btp]    
\begin{center}
\includegraphics[scale=0.3]{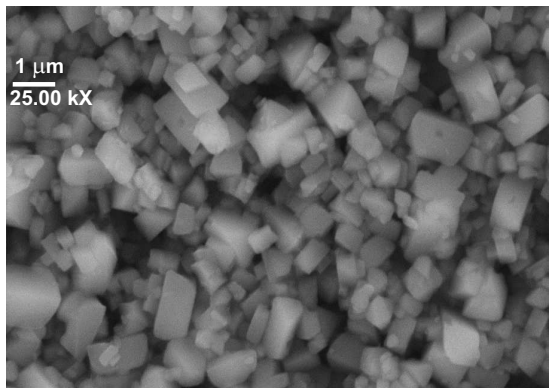}
\end{center} 
\caption{SEM micrographs
of $\protect\beta$ nickel oxalate dihydrate shows prismatic-like shapes.
Average length size (L) and diameter (D) were  1.47 and 0.65 $\protect
\mu$m, respectively.}  
\label{Fig1}  
\end{figure}

\subsection{Structural characterization}

Figure 2a  displays the XRD powder diffraction pattern of $%
NiC_{2}O_{4}\cdot 2H_{2}O$ (blue pattern); it is in good agreement with the
reported (red lines) for the orthorhombic $\beta $-phase of cobalt
oxalate dihydrate (JCPDS file: 25-0250). This figure also shows the Rietveld profile fit (red pattern). In figure 2b
the difference pattern (red) obtained by using the Win-Rietveld software \cite{Cusker99}. 
Background was estimated by linear\- interpolation and
the peak shape was mode\-led by Pseudo-Voigt function. Unit cell parameters
 were $a$ = 11.842 $\mathring{A}$, $b$ = 5.345 $\mathring{A}$, and $c$ =
15.716 $\mathring{A}$ \cite{Deyrieux}. The atomic positions were the
reported by Deyrieux et al., for iron-oxalate \cite{Deyrieux69}; i.e.  $4Co(
\frac{1}{4},\frac{1}{4},0)$, $4Co(0,0,\frac{1}{4})$, $8C_{oxalate}(\frac{1
}{4},\frac{1}{4},0.042)$, $8C_{oxalate}(0,\frac{1}{2},0.042)$, $
16O_{oxalate}(\frac{1}{4},0.0941,0.089)$, $16O_{oxalate}(0,0.691,0.339)$,
$8O_{water}(0.423,\frac{1}{4},0)$, and $8O_{water}(0.173,0,\frac{1}{4})$. The weighted profile and expected residuals
factors obtained in the  refinement were R$_{wp}$ = 32.52, and R$_{exp}$ = 13.52.

\begin{figure}[btp] 
\begin{center}
\includegraphics[scale=0.5]{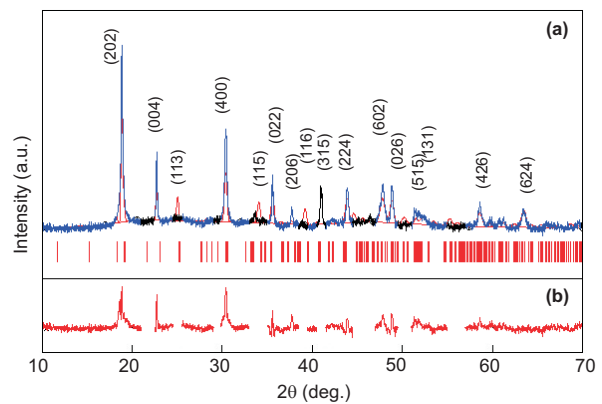} 
\end{center}
\caption{(Color online) (a) X-ray powder
diffraction pattern of nickel oxalate dihydrated (blue), peaks absent (black), Riet\-veld calculated pattern (red) and Miller 
index (red lines) reported for the orthorhombic $Cccm$ phase (JCPDS file: 25 - 0250). (b) Difference pattern (red).}
\label{Fig2}  
\end{figure}

It must be noticed that some low intensity calculated peaks were absent in the experimental pattern (black peaks in figure 2a. 
this could be due to the method of preparation.

Refined cell parameters determined for the orthorhombic phase of Ni-oxalate
sample were $a$ = 11.759(3) $\mathring{A}$, $b$ = 5.327(1) $
\mathring{A}$, and $c$ = 15.654(4) $\mathring{A}$. Schematic
representation of the unit cell is shown in figure 3. There are two
non equivalent positions for nickel ions, designed as Ni1 and Ni2, each
nickel ion is shifted one respect to of other by a translation vector (${\frac{1}{
2},\frac{1}{2},0}$). 

\begin{figure}[btp]   
\begin{center}
\includegraphics[scale=0.4]{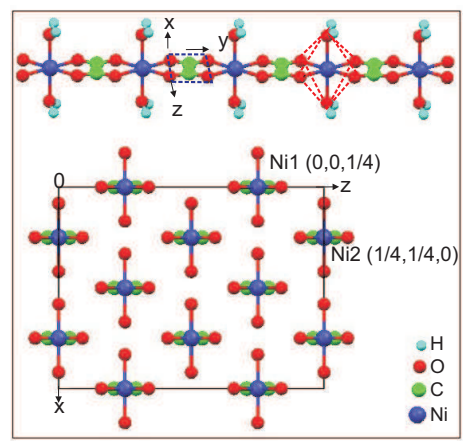}
\end{center}
\caption{(Color online) Ni-oxalate unit
cell of the orthorhombic $\protect\beta -$phase of JCPDS file 25-0250, space
group $Cccm$. The  eight nickel atoms are located in two non-equivalent
positions designed as Ni1 and Ni2.} 
\label{Fig3} 
\end{figure}

Finally, the coherent diffraction size $(D)$ for the sample crystallites was calculated
with the Scherrer equation \cite{Cullity-78} using the $(202)$ peak, the
calculated value was 21.4 nm.

\subsection{Thermal analysis}

Figure 4 shows the TG and DTA curves of $NiC_{2}O_{4}\cdot 2H_{2}O$. 
Two steps of weight loss were observed. The first one ending at 
188 $^{\circ }$C and the second one finishing at 318 $^{\circ }$C. In the first
step the weight loss of 19.93 \% corresponds to the loss of two water
molecules, which agrees with the theoretical value of 19.73 \%. DTA curve
associated with this process shows an endothermic peak at 188 $^{\circ }$C. 
The dehydration reaction is:  
\begin{equation}
NiC_{2}O_{4}\cdot 2H_{2}O_{(s)}\rightarrow NiC_{2}O_{4(s)}+2H_{2}O.
\end{equation}
The weight loss about  36.23 \% in the second step may be associated  to
the decomposition to the anhydrous nickel oxalate to  nickel oxide,
this  agrees with the theoretical value of 39.85 \% \cite{Zhan-05,
Zak-08}. The corresponding DTA curve shows an exothermic peak, at  T = 318 $^{\circ }$C. 
The decomposition reaction in this step can be written as 
\begin{equation}
NiC_{2}O_{4(s)}\rightarrow NiO_{(s)}+CO_{2(g)}+CO_{(g).}
\end{equation} 

\begin{figure}[btp] 
\begin{center}
\includegraphics[scale=0.32]{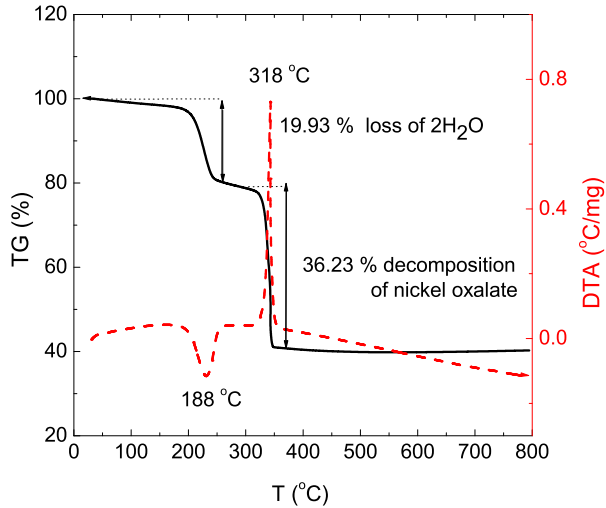} 
\end{center}
\caption{(Color online) The TG (solid
line) / DTA (dash line) curves for $NiC_{2}O_{4}\cdot 2H_{2}O$ at 5 $^{\circ }$C/min heating rate.}   
\label{Fig4}    
\end{figure}

\subsection{Magnetic measurements} 

The molar susceptibility $\chi(T)$, for the sample
is shown in figure 5. It was measured by applying magnetic fields
of 0.01 T, 0.50 T, and 5.00 T in both ZFC, and FC modes. The main panel 
shows two well defined maxima in the susceptibility at 3.3 K and 43 K. We 
also observed changes in the susceptibility at 3.3 K  when the magnetic
field exceeds 5 T, as mentioned in the caption of figure 5.

\begin{figure}[btp]     
\begin{center}
\includegraphics[scale=0.32]{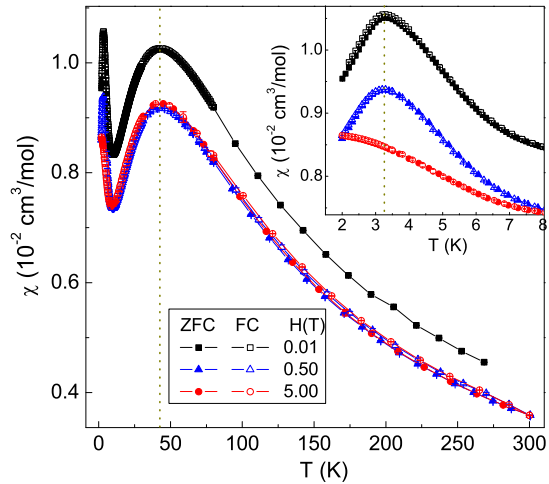}
\end{center}
\caption{(Color online) Main panel: molar
susceptibility of $\protect\beta -NiC_{2}O_{4}\cdot 2H_{2}O$ with maximum
at around 43 K, for both ZFC and FC modes, measured with three applied fields: 0.01 T, 0.50 T and 5.00 T.
Inset: displays the susceptibility at low temperature. The maximum at about 3.3 K changes 
and disappears when the applied magnetic field exceeds 5 T.} 
\label{Fig5}      
\end{figure}

\begin{figure}[btp]     
\begin{center}  
\includegraphics[scale=0.32]{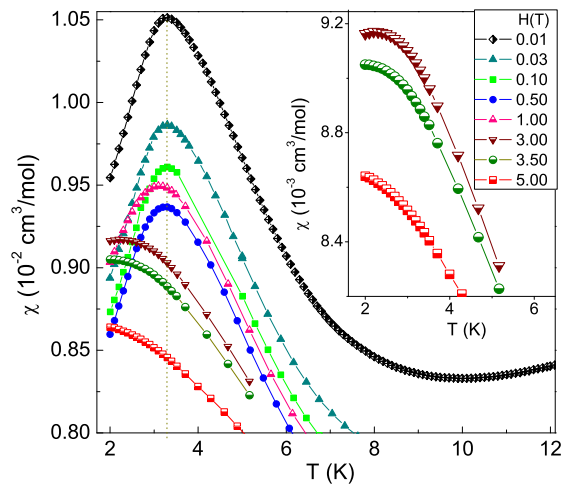} 
\end{center}
\caption{(Color online) This figure presents in the main panel the Molar
susceptibility at low temperature with applied fields from 0.01 to 5.00 T, 
clearly  is seen the evolution of the molar susceptibility with increasing field. 
The Inset displays three curves measured with magnetic fields   
from 3 to 5 T. At a field about  3.5 T  the maximum at 3.3 K dissapears.} 
\label{Fig6}                
\end{figure}

Figure 6 shows susceptibility measurements $\chi(T)$, at 
magnetic fields from 0.01 T to 5 T. Those results indicated than
at a field of 3.5 T, the peak at $T$ = 3.3 K disappears. This change in the magnetic behavior is related with a 
metamagnetic transition. Thus, the  applied field $\geq$
3.5 T, in this compound, is the critical field at which the 
constraints of crystal field are exceeded, and the magnetic behavior
changes.

These results can be compared with those found in other $Ni^{2+}$ systems. 
Evidences of a metamagnetic transition in $Ni(C_{2}O_{4})(bpy)$ was
also obtained in  polycrystalline materials \cite{Yuen-06}. The transition has been
anticipated by slope changes in $M(H)$ with higher applied fields, i.e., $H>$
30 kG and below $T_{N}=26$ K. In this compound, the
one-dimensional magnetic character can be enhanced by replacing the
transaxial aquo group by organic groups, such as bipyridine. This has the
effect of reducing the interchains interaction and consequently the N\'{e}el temperature, 
and the critical fields required for the meta\-magnetic
transitions \cite{Oti-95, Mao-96}. Important examples about these behavior are $
CoCl_{2}\cdot 2H_{2}O$, and $\alpha -Co(pyridine)_{2}Cl_{2}$ \cite
{Fon-78}. The details of the magnetic properties of these mate\-rials are best
understood by examining single crystal data because the transitions are
sensitive to the orientation of the applied field  \cite{Keffer-73}.

It is important to mention that in figure 5, we show that the magnetic
susceptibility presents a wide  maximum at about 43 K. From 100 K to room temperature 
$\chi(T)$ smoothly decreases, and a paramagnetic Curie-Weiss behavior  can be fitted.
The maximum value at 43 K changes with the 
magnetic field applied as shown in this figure. The changes 
are on the range from 0.91x$10^{-2}$ to 1.03x$10^{-2}$ $
cm^{3}/mol$. This behavior is quite  similar to the reported in other compounds, as $
Ni(pip)(C_{2}O_{4})$, where pip= piperazine, which is formed by chains of 
$[Ni(ox)]_{n}$ and $[Ni(pip)]_{n}$. In that example, a broad maximum 
is observed and occurs at 53(1) K \cite{Keene-04}. 
So, accordingly the broad maximum in the susceptibility in our compound is a typical 
behavior of a low dimensional antiferromagnetic system \cite{Bonner-64}.

In figure 7 we  show a plot of the inverse of susceptibility, $
\chi(T)^{-1}$  at $H$ = 0.01, 0.50 and 5.00 T, for $
\beta -NiC_{2}O_{4}\cdot 2H_{2}O$. The analysis of these measurements was
performed by fitting a Curie-Weiss from room temperature to 100 K. The fit
parameters were Weiss temperature $\theta _{\omega}$ and Curie constant $C$
varying from -76.90 to -99.82 K and from 1.35 to 1.67 cm$^{3}$mol$^{-1}$K,
respectively.   

The fit parameters could be used to calculate the effec\-tive magnetic moment $
\mu _{eff}$ per mole from equation $\mu _{eff}$= 2.84$[C]^{1/2}$= 2.84 $[\chi T]^{1/2}$, which is shown in figure 8.

\begin{figure}[btp]         
\begin{center}
\includegraphics[scale=0.32]{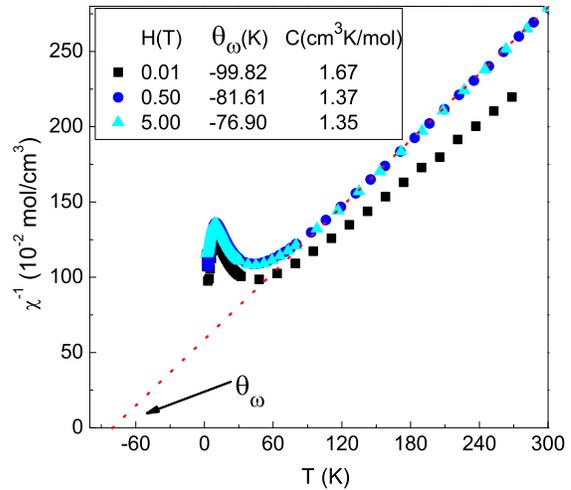}
\end{center}
\caption{(Color online) Inverse
susceptibility at different magnetic fields, $\protect\chi ^{-1}(T)$ for dihydrate nickel
oxalate. The Weiss temperature $\protect\theta _{\omega}$, changes from about -76.90
to -99.82 K. The Curie constant also changes from 1.35 to 1.67 cm$^{3}$mol$
^{-1}$K, from high to low magnetic fields.} 
\label{Fig7}      
\end{figure}

\begin{figure}[btp]                  
\begin{center}
\includegraphics[scale=0.32]{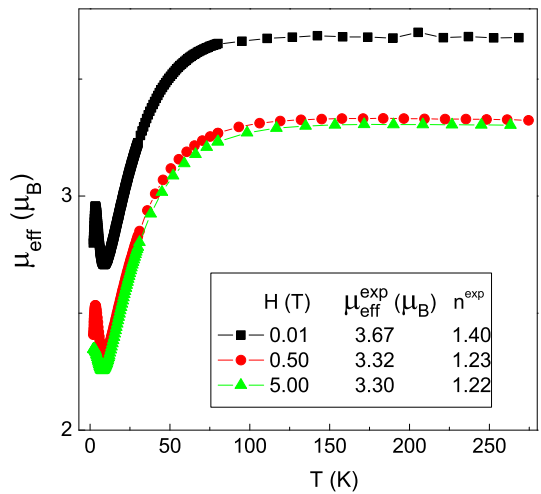}
\end{center}
\caption{(Color online) Effective Bohr
magneton moments $\protect\mu _{eff}$\ per mol as a function of temperature.
At room tempe\-rature the values change from 3.30 to about 3.67 $\protect\mu_{B}$, 
depen\-ding on the magnetic field applied. $n^{exp}$ is the number of unpaired electrons.  } 
\label{Fig8}               
\end{figure}    

An important aspect of the behavior $\mu _{eff}$, for this compound, and  related to Curie constant, and the  
large ne\-gative Weiss constant,  is  indicative of a
significant antiferromagnetic exchange interactions between neighbouring
nickel ions, and  a big  degree of frustration \cite{Ramirez-01}. It is quite possible because the metamagnetic transition obtained 
in figures 5 and 6. As observed,  this dramatic change of the $\mu _{eff}$ is the effect of the change of the Curie constant, and then 
via the number of unpaired electrons. Thus,  accordingly to $\mu _{eff}$= $g[n(n+1)]^{1/2}$, the number of unpaired
electrons $n$ in $NiC_{2}O_{4}\cdot 2H_{2}O$ can be calculated  as 1.40, 1.23 and 1.22  respectively for applied field of 0.01 T, 0.50 T and 5.0 T. 

In order to understand more about the magnetic cha\-racteristics of this
oxalate, we studied the $M-H$ isothermal measurements from  the temperature range
2 - 80 K (see figure 9), from these data we extracted the 
coercive field, which shows a small  measurable and perceptible exchange bias. Our measurements of the coercive field, although small, 
were carefully checked and are below of the possible errors. The results are 
displayed in figure 10.

\begin{figure}[btp]  
\begin{center} 
\includegraphics[scale=0.32]{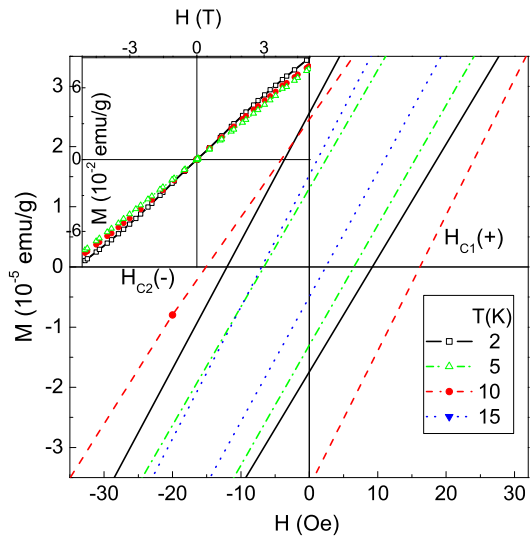}      
\end{center}
\caption{(Color online) Isothermal
magnetization measurements $M-H$ from 2 to 15 K shown in the main panel. 
Note the coercive field and the small exchange bias.
Inset shows the $M(H)$ measurements at 4 T and different
temperatures from 2 to 80 K. At low fields (main panel) the hysteric effect
is clearly observed. The asymmetric behavior in the coercive field is related to a exchange
bias effect due to canted spins, but
other type of interaction is not discarted as driven by
Dzyalochinsky-Moriya type exchange (DM).} 
\label{Fig9}   
\end{figure}   

\begin{figure}[btp] 
\begin{center}
\includegraphics[scale=0.3]{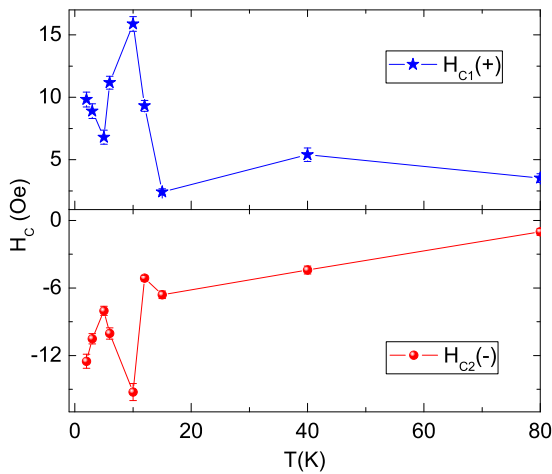}
\end{center}
\caption{(Color online) Coercive Field
vs Temperature determined by isothermal magnetic measurements.} 
\label{Fig10}      
\end{figure}

This exchange bias can be explained as the effect of inter and intra chains
interactions in this compound. The interaction of metallic ions between
chains is at the origin of spin canting. This small but measurable exchange
bias indicates that instead of produce a pure antiferromagnetic order, the
magnetic order has been distorted by canted spins giving rise to weak ferromagnetism. Since weak ferromagnets are
 a delicate balance of opposing forces, it is not surprising to find that many are also metamagnets \cite{Stry-77}.  
  
Experimentally, it is important to mention that great care was taken when
measuring the exchange bias. Our SQUID magnetometer is provided with a Mu
metal shielding. At the moment of performing the magnetization measurements
a flux gate magnetometer was used to demagnetize the superconducting coil.
This procedure reduces the magnetic field to a very small value to about
0.001 G or less and the Mu shielding eliminates external magnetic
influences, as the earth magnetic field.

\section{Conclusions}       

Microcrystals single phase of orthorhombic nickel (II) oxalate dihydrate were prepared by
soft solution che\-mistry, as observed by XRD powder diffraction. Chemi\-cal analysis, DTA and TG studies revealed that
the microcrystals are of high purity. $\chi(T)$ measurements showed the
existence of two maxima at 3.3 K and at 43 K. The first one at low temperature changes and disappears with applied magnetic field,
this change is due to meta\-magnetic transition, the maximum disappears with applied field about $
\geq$ 3.5 T. The second maximum indicates an
antiferromagnetic order, with interactions due to coupled chains via inter- and intrachain interactions 
and/or DM type-exchange. The effects of 
interchain interactions disturb the AF coupling, distorting it and canting
spins, which in turn produces a weak ferromagnetic order. This WF is
evident by hysteresis measurements in M-H isothermal measurements.

\begin{acknowledgments}
Partial support for this work is gratefully acknow\-ledge to CONACyT, project No.44296/A-1 and Scho\-larship CONACyT, register No. 
188436 for E. Romero; VIEPBUAP, project No. MEAM-EXC10-G. RE, thanks to CONACyT Project 129293(Ciencia B\'{a}sica), DGAPA-UNAM project No. 
IN100711, project BISNANO 2011, and project PICCO 11-7 by The Departament of Distrito Federal, M\'{e}xico.
\end{acknowledgments}

\end{document}